\documentclass[twocolumn,final,times,sort&compress]{elsarticle}
\usepackage{graphicx,amssymb,amsmath}
\journal{Physica B}

\begin{document}

\begin{frontmatter}

\title{The Schottky-type specific heat as an indicator of relative degeneracy between ground and first-excited states: the case study of regular Ising polyhedra\tnoteref{grant}}
\tnotetext[grant]{This work was financially supported by Ministry of Education of Slovak Republic provided under the VEGA grant No. 1/0043/16 
and by the Slovak Research and Development Agency under the contract No. APVV-14-0073.}
\author[UFV]{Katar\'ina Kar\v{l}ov\'a}
\author[UFV]{Jozef Stre\v{c}ka\corref{coraut}} 
\cortext[coraut]{Corresponding author}
\ead{jozef.strecka@upjs.sk}
\author[UMV]{Tom\'a\v{s} Madaras}
\address[UFV]{Institute of Physics, Faculty of Science, P. J. \v{S}af\'{a}rik University, Park Angelinum 9, 040 01 Ko\v{s}ice, Slovak Republic}
\address[UMV]{Institute of Mathematics, Faculty of Science, P. J. \v{S}af\'{a}rik University, Jesenn\'a 5, 040 01 Ko\v{s}ice, Slovak Republic}

\begin{abstract}
The specific heat of regular Ising polyhedra is investigated in detail as a function of temperature and magnetic field. It is shown that the regular Ising polyhedra display diverse double-peak temperature dependences of the specific heat whenever the magnetic field approaches a level-crossing field.  The Schottky theory of a two-level system often provides a plausible explanation of a height and position of low-temperature peak, which emerges in the specific heat of a regular Ising polyhedron due to low-lying excitations from a ground state to a first-excited state. The height and position of Schottky-type maximum depends essentially on a relative degeneracy of the ground state and first-excited state, which are in general quite distinct in geometrically frustrated Ising spin clusters. Low-temperature variations of the specific heat with the magnetic field exhibit multipeak structure with two peaks (of generally different height) symmetrically placed around each level-crossing field. 
\end{abstract}


\end{frontmatter}

\section{Introduction}

The specific heat of a substance is defined as the amount of heat, which is needed to raise the temperature by a unit degree. The specific heat of solid-state materials usually exhibits an outstanding temperature dependence, because the overall thermal behavior comes from an interplay between several specific contributions such as lattice vibrations, electronic distributions and magnetic excitations. It is worthwhile to remark that the magnetic contribution to the specific heat spreads just over a certain range of temperatures in contrast to the lattice and electronic contributions, which are also relevant at higher temperatures where the lattice part usually prevails over all other contributions. Although the specific heat is just a mean quantity reflecting the overall energy spectrum, under certain conditions it may bring insight into microscopic details such as degeneracy of low-lying energy levels. To illustrate the case, we will perform in the present work a comprehensive analysis of the magnetic part of the specific heat of the regular Ising polyhedra in an external magnetic field.  

The magnetic part of the specific heat (hereafter simply called as the specific heat) can be often approximated in a certain temperature range by the Schottky theory of a two-level system \cite{gopa66,path96}. The associated round maximum of the specific heat, the so-called Schottky anomaly or Schottky peak, has been experimentally detected in various magnetic compounds among which one could mention the molecular magnets Fe$_6$, Fe$_{10}$ \cite{affr99} and Fe$_{12}$ \cite{affr02}, heavy-fermion compounds  Fe$_2$VAl \cite{luec99}, PrOs$_4$Sb$_{12}$ \cite{tsuj04} and PrAg$_2$In \cite{sato14}, pyrochlore spin-ice compounds Dy$_2$Ti$_2$O$_7$ \cite{higa02} and Tb$_2$Ti$_2$O$_7$ \cite{hama04}, or several magnetic insulators BaCuO$_2$ \cite{fish00}, TbFe$_3$(BO$_3$)$_4$ \cite{popo07}, Cu$_7$Cl$_2$(OH)$_6$(pen-disulfide)$_3$ \cite{toko10} and NH$_4$CuPO$_4$ \cite{chak15}. The geometrically frustrated spin systems such as the spin-tube compound [(CuCl$_2$tachH)$_3$Cl]Cl$_2$ \cite{ivan10} or the Shastry-Sutherland-lattice compounds R$_2$BaPdO$_5$ (R=Nd,Eu,Gd and Ho) \cite{ozaw05} also display the pronounced Schottky-type anomaly in the specific heat. 

Recently, it has been theoretically verified that quantum Heisenberg spin clusters can also exhibit the marked Schottky maxima due to a mutual competition between antiferromagnetic interactions and magnetic field \cite{park00,kons05,efre06,huch11}. The essential ingredient for observing the Schottky anomaly in the specific heat of antiferromagnetic Heisenberg spin clusters are abrupt magnetization jumps, which appear in a stepwise magnetization curve including one or more intermediate plateaus due to a crossing of energy levels \cite{schn09a,schn09b,hone09,schn10}. In our preceding work we have convincingly evidenced an existence of intermediate plateaus and magnetization jumps in a magnetization process of the regular Ising polyhedra with a unique antiferromagnetic nearest-neighbor interaction \cite{stre15}. In the present follow-up work, we will examine the specific heat of the regular Ising polyhedra, which seems not to be dealt with previously.  

The paper is organized as follows. In Section \ref{SH} we briefly describe the method used for a calculation of the specific heat of the regular Ising polyhedra. Section \ref{method} deals with the Schottky theory, which often enables to determine the height and position of specific-heat maximum. Section \ref{result} is devoted to a discussion of the most interesting results for temperature and field dependences of the specific heat. Finally, Section \ref{conclusion} provides a brief summary of the most important findings.

\section{Specific heat of the regular Ising polyhedra}
\label{SH}
In the present work we will examine in detail temperature and magnetic-field dependences of the specific heat of the regular Ising polyhedra (tetrahedron, octahedron, cube, icosahedron, dodecahedron) with the uniform antiferromagnetic interaction between nearest-neighbor spins, which are defined through the following Hamiltonian
\begin{equation}
{\cal H} = J \sum_{\langle i,j \rangle}^{N_b} S_i S_j - h \sum_{i=1}^{N} S_i. 
\label{ham}
\end{equation} 
In above, $S_i = \pm 1$ represents the Ising spin placed at $i$th vertex of a regular polyhedron, the first summation takes into account the antiferromagnetic Ising interaction $J>0$ between nearest-neighbor spins, the second summation accounts for the Zeeman's energy of individual magnetic moments in an external magnetic field $h>0$ and finally $N$ $(N_b)$ stands for the total number of spins (interactions) within a given regular polyhedron. The magnetization curves and magnetocaloric properties of the regular Ising polyhedra defined through the Hamiltonian (\ref{ham}) were thoroughly studied in our previous work \cite{stre15}, which includes exact results for the partition function, Gibbs free energy $g$ per one spin, magnetization and entropy of all regular Ising polyhedra. In this work we will  therefore limit our particular attention to a detailed examination of the specific heat of the regular Ising polyhedra, which can be straightforwardly calculated from the exact results (6)-(10) presented in Ref.~\cite{stre15} for the Gibbs free energy according to the basic thermodynamic relation
\begin{eqnarray}
C_{h} = -T \left( \frac{\partial^2 g}{\partial T^2} \right)_{h}.
\label{C_{h}}
\end{eqnarray}
\section{Schottky theory}
\label{method}

It has been convincingly evidenced in our previous work \cite{stre15} that the regular Ising polyhedra with a unique antiferromagnetic nearest-neighbor coupling exhibit magnetization curves with notable magnetization jumps, which arise out from a crossing of the lowest-energy levels. It could be expected that the specific heat of the regular Ising polyhedra can be often approximated at low-enough temperatures by the Schottky theory of a two-level system.
 
\begin{figure}[t]
\begin{center}
\vspace{-0.1cm}
\includegraphics[width=0.39\textwidth]{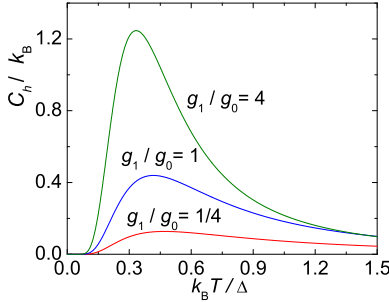}
\vspace*{-0.5cm}
\caption{The specific heat of a two-level system as a function of temperature for three different values of a relative degeneracy $g_{1} / g_{0}$.}
\label{fig1}
\end{center}
\end{figure}

Let us consider a two-level system  composed of two different energy levels: the ground-state energy level with energy $E_{0} = 0$ and the excited energy level $E_{1} = \Delta > 0$. The parameter $\Delta$ simultaneously determines a splitting of both  energy levels $\Delta = E_{1} - E_{0} > 0$. The partition function of a two-level system can be calculated from
\begin{equation}
 Z = g_{0}\exp\left(-\beta E_{0}\right) + g_{1}\exp\left(-\beta E_{1}\right)  
   = g_{0} + g_{1}\exp\left(-\beta\Delta\right),
	\label{ZF}
\end{equation}
where $g_{0}$ denotes degeneracy of a ground state and $g_{1}$ is respective degeneracy of an excited state. The relevant enthalpy is given by
\begin{equation}
H = -\frac{\partial \ln {Z}}{\partial \beta}= -\frac{1}{Z}\frac{\partial {Z}}{\partial \beta}.
\label{ent}
\end{equation}
Substituting the partition function (\ref{ZF}) to a definition of the enthalpy (Eq.~(\ref{ent})) leads after a few algebraic manipulations to the following relation 
\begin{eqnarray}
H = \Delta \frac{g_{1}}{g_{0}\exp\left(\beta \Delta\right) + g_{1}}.
\label{enth}
 \end{eqnarray}
The specific heat at a constant external field $h$ can be obtained by differentiation of the enthalpy (\ref{enth}) with respect to the temperature $C_{h} = (\frac{\partial H}{\partial T})_{h}$, which gives the formula
\begin{eqnarray}
\frac{C_{h}}{k_{\rm B}}= \frac{\frac{g_{1}}{g_{0}}\left(\beta \Delta\right)^2}{\left[\exp\left(\frac{\beta \Delta}{2}\right)+\frac{g_{1}}{g_{0}}\exp\left(-\frac{\beta \Delta}{2}\right)\right]^2}.
\label{c/k}
 \end{eqnarray}
Typical temperature dependences of the specific heat (\ref{c/k}) are plotted in  Fig.~\ref{fig1} for three different values of a relative degeneracy $g_{1}/g_{0}$. As one can see, the specific heat shows round Schottky-type maximum, whereas
the peak position shifts to lower temperatures and increases its height when the relative degeneracy $g_{1}/g_{0}$ increases. As a result, the higher but narrower Schottky-type maximum is observed for greater values of the relative degeneracy $g_{1}/g_{0}$.

Consider at first a simple case, when the ground-state degeneracy equals to the degeneracy of excited state $g_{1}=g_{0}=g$. The expression (\ref{c/k}) then simplifies to the usual expression for the Schottky anomaly \cite{path96} given by
\begin{eqnarray}
\frac{C_{h}}{k_{\rm B}}=\frac{\left(\frac{\beta \Delta}{2}\right)^2}{\cosh^2\left(\frac{\beta \Delta}{2}\right)}. 
\label{shotky}
 \end{eqnarray}
The temperature dependence given by Eq.~(\ref{shotky}) is characterized by a maximum, which can be obtained from a  stationary point condition  $\left(\frac{\partial {C_{h}}}{\partial T} \right)_{h}= 0$. This condition is equivalent to solving the following transcendental equation
\begin{eqnarray}
\frac{\beta_{max} \Delta}{2}=\coth\left(\frac{\beta_{max} \Delta}{2}\right).
\label{trans}
 \end{eqnarray}
The substituting the extreme condition (\ref{trans}) to Eq.~(\ref{shotky}) then determines the height of the Schottky-type maximum 
\begin{eqnarray}
\frac{C_{max}}{k_{\rm B}}=\frac{1}{\sinh^2\left(\frac{\beta_{max} \Delta}{2}\right)}.
\label{Cmax1}
 \end{eqnarray}

For a more general case with unequal degeneracy of ground and excited states ($g_{0} \neq g_{1}$), the same procedure leads to the following transcendental equation
\begin{eqnarray}
\frac{\beta_{max} \Delta}{2}=\frac{\exp\left(\frac{\beta_{max} \Delta}{2}\right) + \frac{g_{1}}{g_{0}}\exp\left(-\frac{\beta_{max} \Delta}{2}\right)}{\exp\left(\frac{\beta_{max} \Delta}{2}\right) - \frac{g_{1}}{g_{0}}\exp\left(-\frac{\beta_{max} \Delta}{2}\right)},
\label{transcen}
 \end{eqnarray}
which determines a locus of the Schottky-type maximum \cite{gopa66}. The substitution of the extremum condition (\ref{transcen}) to Eq.~(\ref{c/k}) determines the height of Schottky-type maximum
\begin{eqnarray}
\frac{C_{max}}{k_{\rm B}}=4 \frac{g_{1}}{g_{0}}\left[\exp\left(\frac{\beta_{max} \Delta}{2}\right)-\frac{g_{1}}{g_{0}}\exp\left(-\frac{\beta_{max} \Delta}{2}\right)\right]^{-2}.
\label{Cmax}
 \end{eqnarray}
\begin{figure*}
\begin{center}
\vspace{-2.0cm}
\hspace{-1.0cm}
\includegraphics[width=0.44\textwidth]{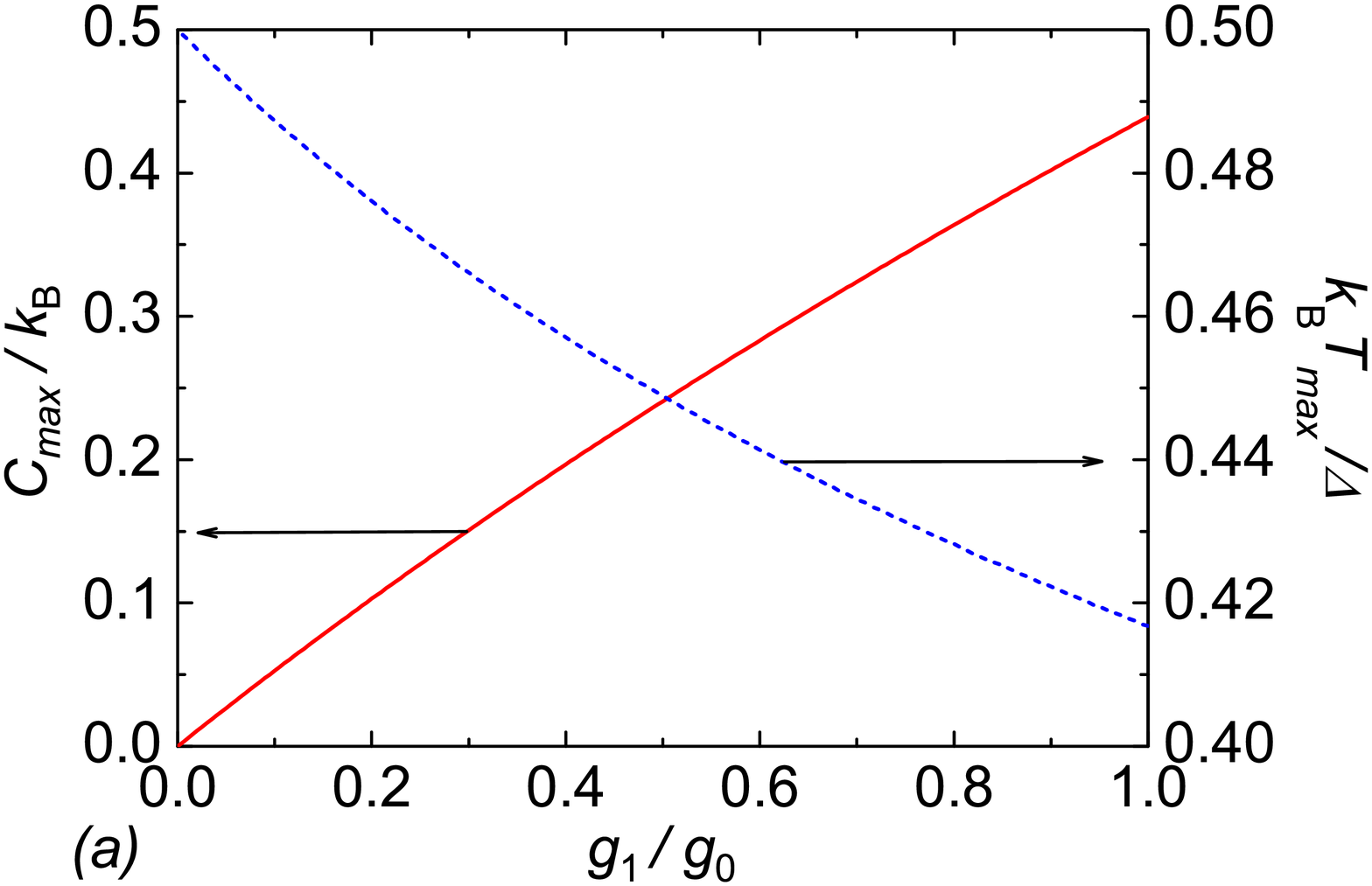}
\hspace{-0.2cm}
\includegraphics[width=0.44\textwidth]{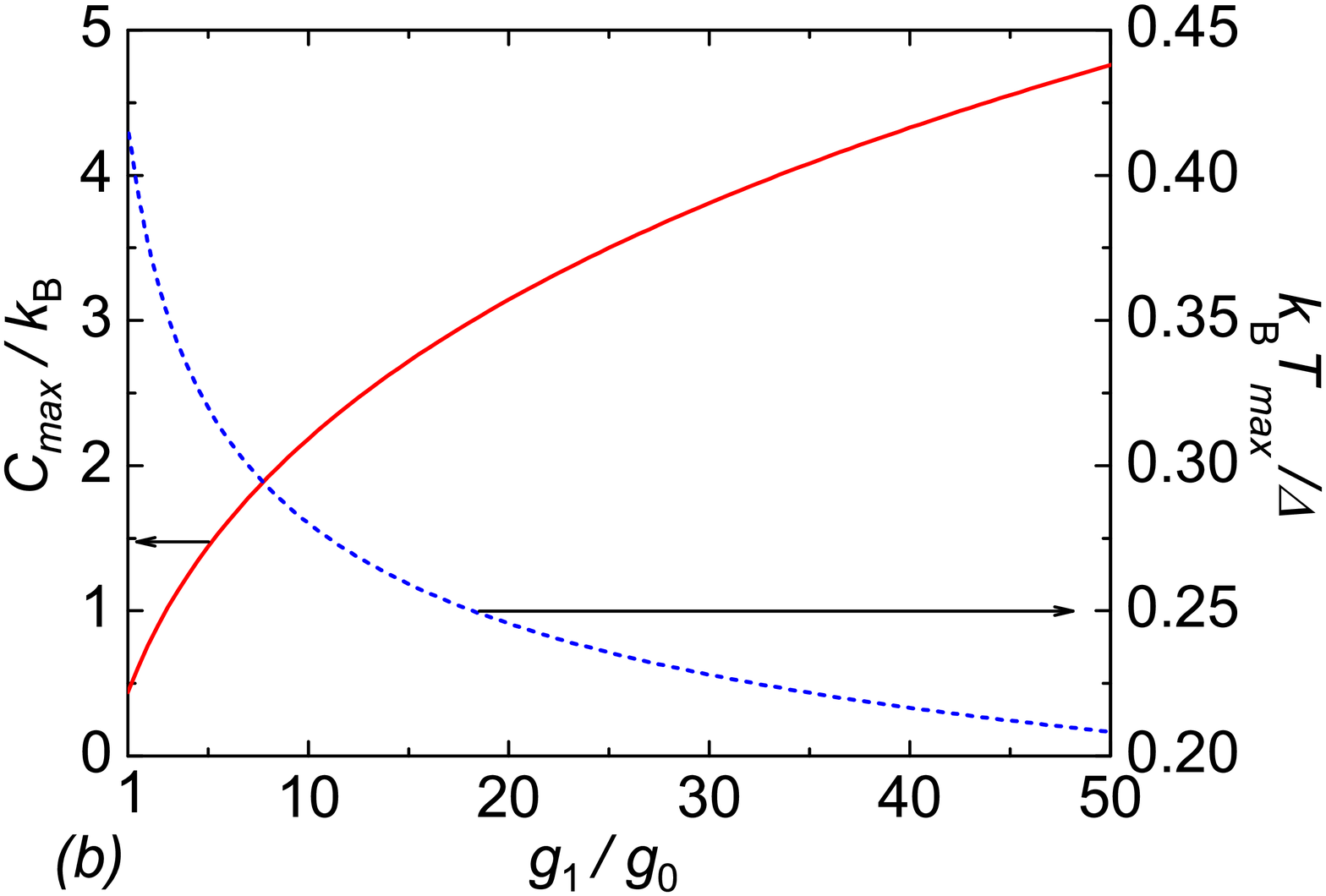}
\hspace{-1.0cm}
\vspace{-0.5cm}
\caption{The height and position of the Schottky-type maximum as a function of the relative degeneracy within the range: (a) $g_{1} / g_{0} \in (0;1)$; (b) $g_{1} / g_{0} \in (1;50)$. Note that the height is scaled with respect to a left axis, while the position is scaled with respect to a right axis.}
\label{fig:vap}
\end{center}
\end{figure*}
It can be readily seen from  Eqs.~(\ref{transcen}) and (\ref{Cmax}) that the position and height of Schottky-type maximum depends just on the relative degeneracy $g_{1}/g_{0}$ between the excited state and ground state. By solving of the transcendental Eq.~(\ref{transcen}) using bisection method, it is possible to find a position and subsequently from Eq.~(\ref{Cmax}) also a height of the Schottky-type maximum, which are displayed in Fig. \ref{fig:vap} as a function of the relative degeneracy $g_{1}/g_{0}$. In accordance with the plots presented in Fig.~\ref{fig:vap}, one may generally conclude that the height of Schottky-type maximum monotonically increases with increasing the ratio $g_{1} / g_{0}$, while the temperature corresponding to the Schottky-type maximum simultaneously decreases. The numerical values of the height and position of the Schottky-type maximum are listed in Tab.~\ref{piky} for a few selected values of the relative degeneracy $g_{1}/g_{0}$.  
\begin{table}
\vspace{-1.5cm}
\caption{The position and height of the Schottky-type maximum for a few selected values of the relative degeneracy $g_{1}/g_{0}$ between the first excited state and the ground state.}
\begin{center}
\begin{tabular}{|c|c|c|}
\hline  $g_{1}/g_{0}$ & $k_{\rm B}T_{max}/ \Delta$  & $C_{max} / k_{\rm B}$\\ 
\hline  1/4 & 0.471 & 0.127\\ 
\hline  1/2 & 0.449 & 0.241\\
\hline  2/3 & 0.437 & 0.311\\
\hline  1 & 0.417 & 0.439\\
\hline  3/2 & 0.394 & 0.610\\    
\hline  5/3 & 0.388 & 0.663\\ 
\hline  9/5 & 0.383 & 0.703\\  
\hline  2 & 0.377 & 0.762\\
\hline  3 & 0.352 & 1.023\\
\hline  4 & 0.334 & 1.246\\
\hline  6 & 0.309 & 1.617\\
\hline  8 & 0.292 & 1.923\\
\hline  12 & 0.271 & 2.416\\     
\hline  20 & 0.246 & 3.144\\  
\hline  48 & 0.210 & 4.678\\
\hline  64 & 0.200 & 5.263\\    
\hline 
\end{tabular}
\end{center}
\label{piky}
\end{table}

\section{Results and discussion}
\label{result}
In this section, let us proceed to a discussion of the most interesting results for the specific heat of regular Ising polyhedra at a constant magnetic field $h$. For simplicity, our further discussion will be restricted to a particular case with the antiferromagnetic Ising interaction $J>0$, which causes a geometric spin frustration in all spin clusters involving faces with odd number of spins. It is noteworthy that the geometric spin frustration of the regular Ising polyhedra causes a rather high degeneracy of the most of energy levels, see Ref.~\cite{stre15} for a microscopic description of spin arrangements pertinent to individual energy levels. 

\begin{figure*}[t]
\begin{center}
\vspace{-0.5cm}
\centering
\hspace{-1.4cm}
\includegraphics[width=0.4\textwidth]{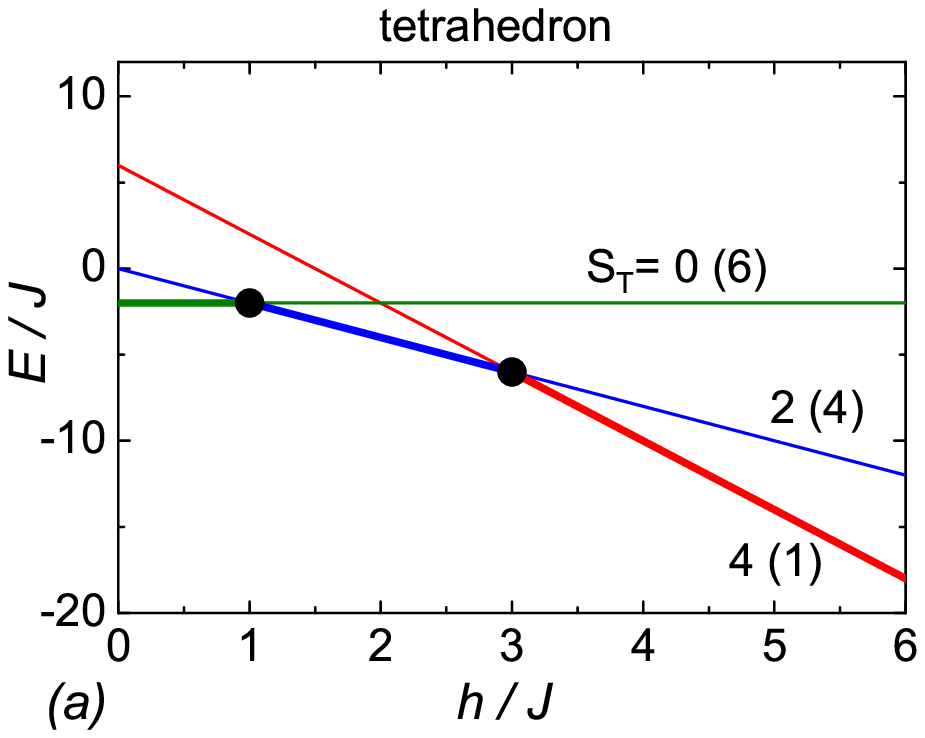}
\hspace{-1.2cm}
\includegraphics[width=0.4\textwidth]{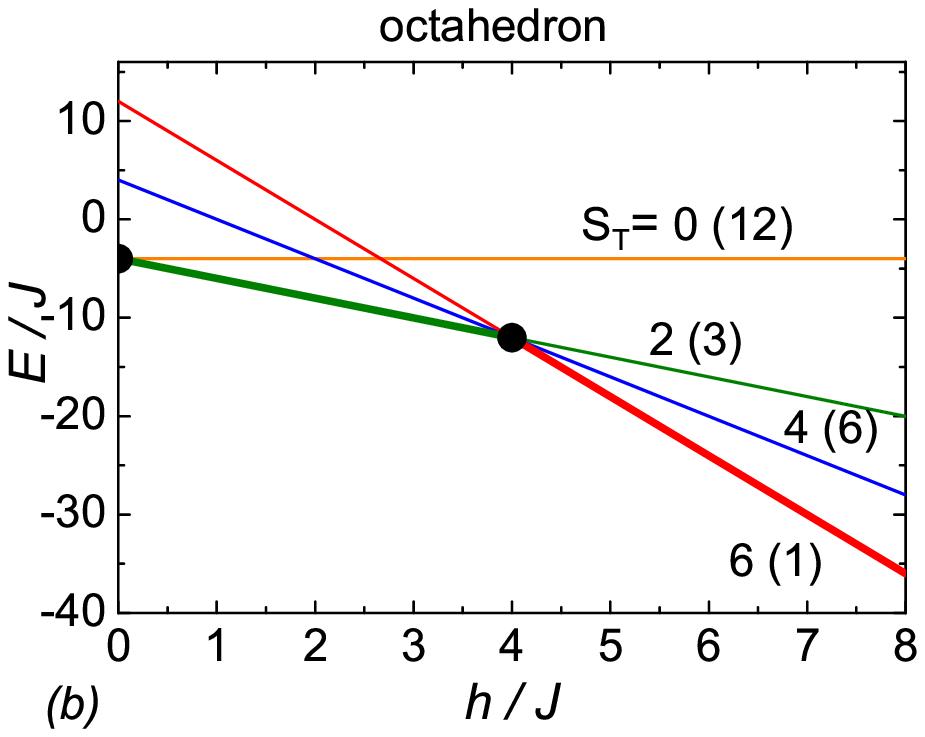}
\hspace{-1.2cm}
\includegraphics[width=0.4\textwidth]{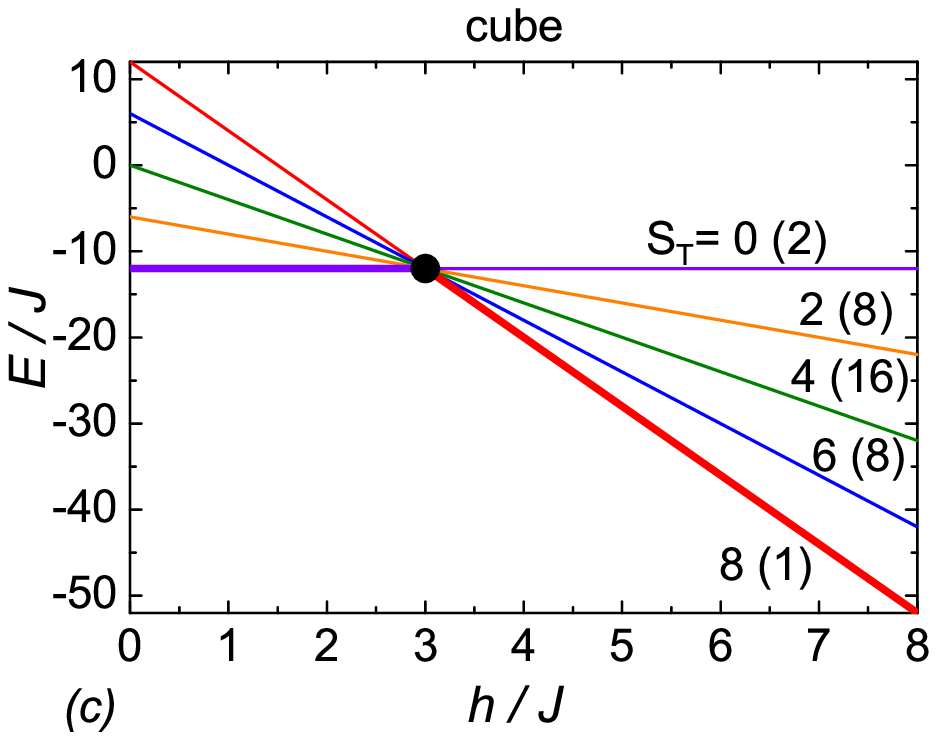}
\hspace{-2.1cm}
\vspace{-0.5cm}
\hspace{-2.5cm}
\includegraphics[width=0.4\textwidth]{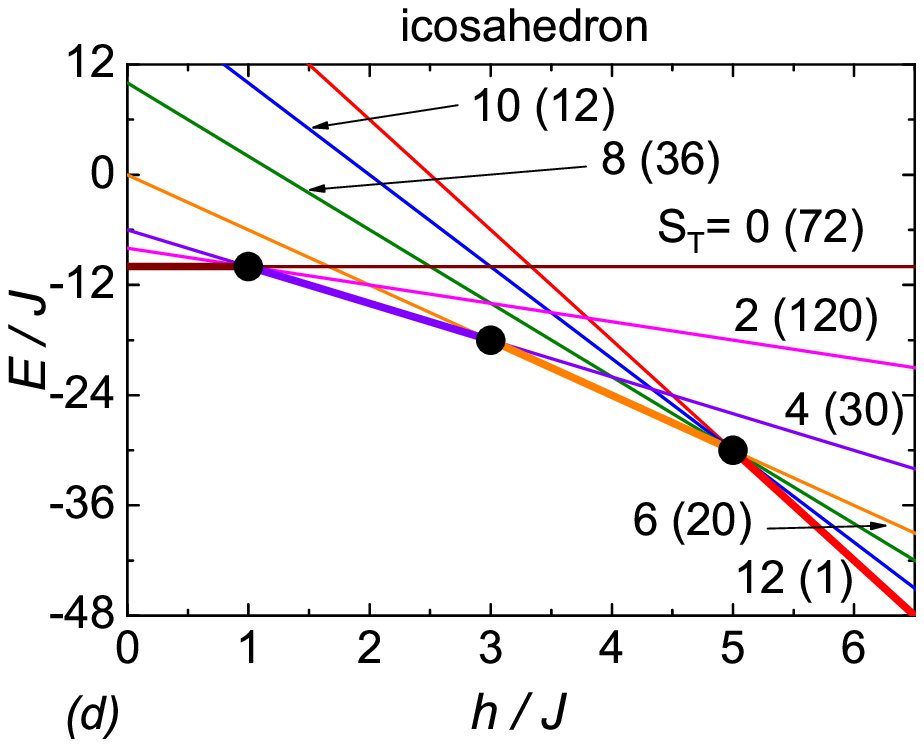}
\hspace{-0.6cm}
\includegraphics[width=0.4\textwidth]{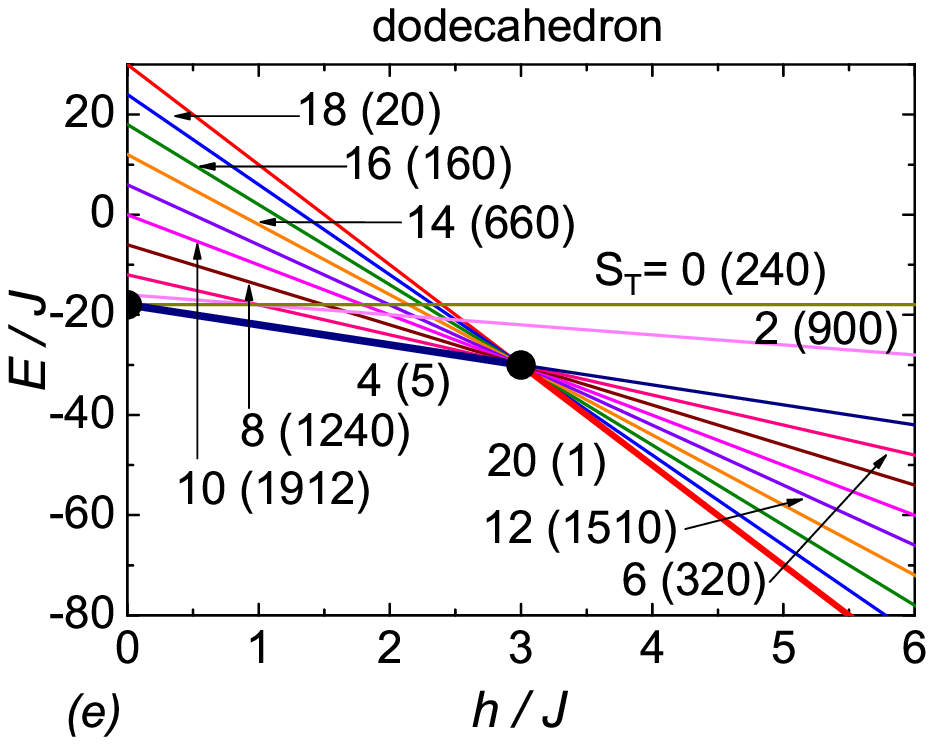}
\vspace{-0.5cm}
\caption{Energy spectra of regular Ising polyhedra involving just lowest-energy levels from all sets with non-negative total spin $S_{T} \geq 0$: (a) tetrahedron, (b) octahedron, (c) cube, (d) icosahedron, (d) dodecahedron. Thick lines highlight energy levels, which correspond to a ground state in a given field range. Numbers quoted in round brackets determine a degeneracy of a given energy level.}
\label{fig:spectrum}
\end{center}
\end{figure*}

In general, the specific heat is fundamentally influenced by the overall energy spectrum, so let us start our discussion with an overview of energy spectra of the regular Ising polyhedra. The dependence of energy levels on the external magnetic field, which is plotted in Fig.~\ref{fig:spectrum} for all five regular Ising polyhedra, implies that an increase of the magnetic field may  give rise to a few crossing of energy levels. The magnetic fields, at which level crossing occurs, correspond to critical fields related to magnetization jumps comprehensively studied in our previous work \cite{stre15}. If only two energy levels cross each other at the particular critical field, then, the specific heat could be satisfactorily described at low enough temperatures by the Schottky theory presented in Section \ref{method}. 

The emerging peaks of the specific heat at low enough temperatures originate from low-lying excitations between the ground and first-excited states, whereas one should generally resort to the generalized Schottky theory valid for arbitrary relative degeneracy $g_{1}/g_{0}$ with respect to a miscellaneous degeneracy of most of the energy levels. The difference between energy levels is nearby critical fields proportional to a deviation of the magnetic field from its critical value: $\Delta = \delta S_{T} (h-h_{c})$, where $\delta S_{T} = S_{T}^{(0)} - S_{T}^{(1)} $ denotes a respective  difference between the total spin of the ground state and first-excited state. The respective difference between the total spin in the most cases equals to $|\delta S_{T}| = 2$, because a crossing of the energy levels from two adjacent sectors with the total spin $S_{T}$ and $S_{T} + 2$ is the most common (the latter sector is obtained from the former one by inverting one spin). In the following, our further attention will be focused on temperature and field dependences of the specific heat. The main emphasis is laid on temperature dependences of the specific heat at magnetic fields in a vicinity (slightly above or below) of level-crossing fields and the magnetic-field dependences of the specific heat at low enough temperatures. 
\begin{figure*}[t]
\begin{center}
\hspace{-1.4cm}
\includegraphics[width=0.39\textwidth]{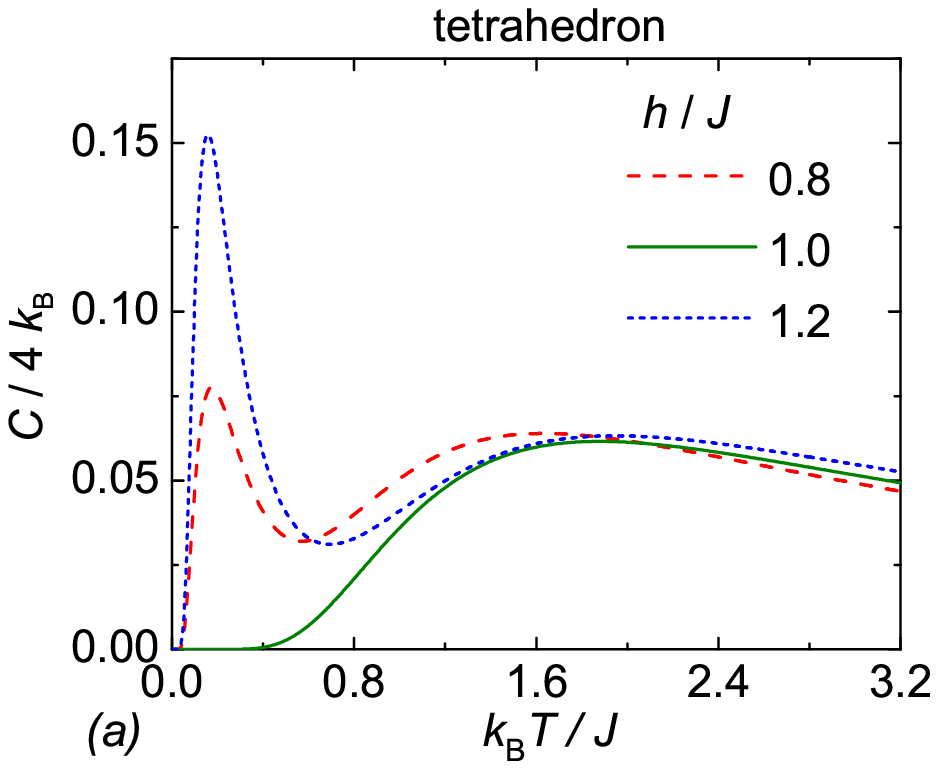}
\hspace{-1cm}
\includegraphics[width=0.39\textwidth]{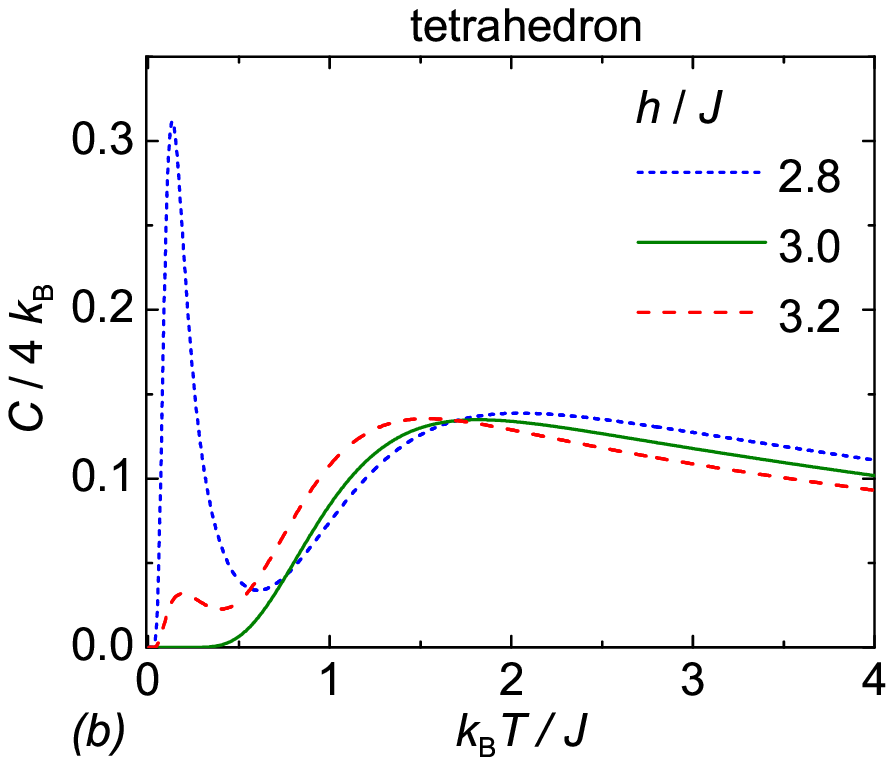}
\hspace{-1cm}
\includegraphics[width=0.39\textwidth]{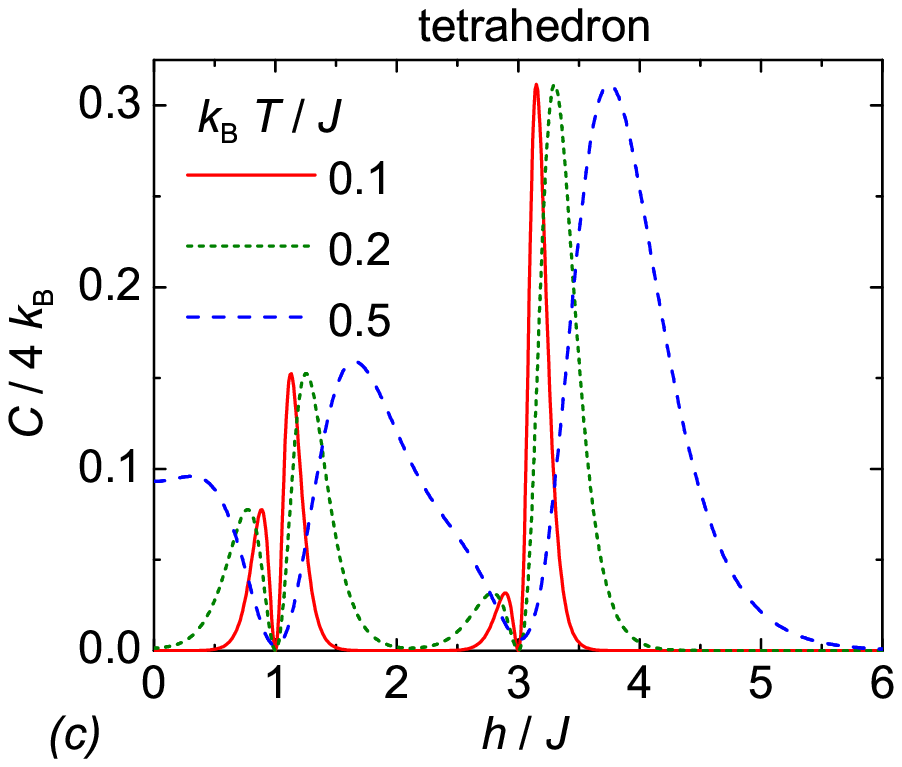}
\vspace{-0.5cm}
\hspace{-1.4cm}
\caption{(a)-(b) Temperature dependences of the specific heat of the Ising tetrahedron at magnetic fields close to the critical fields $h_{c1}=1$ and $h_{c2}=3$; (c) the magnetic-field dependence of the specific heat of the Ising tetrahedron at different values of temperature.}
\label{fig:tetrakap}
\label{tetra}
\end{center}
\end{figure*}

The Ising tetrahedron exhibits in a vicinity of the first critical field $h_{c1}/J = 1$ low-temperature peaks connected with thermal excitations between the lowest-energy level from the set with the total spin $S_{T}=0$ and the lowest-energy level from the set with the total spin $S_{T}=2$ [see Fig.~\ref{fig:tetrakap} (a)]. The readers interested in details concerned with microscopic spin arrangements relevant to particular spin configurations are referred to our previous work \cite{stre15}. The relative degeneracy between the first excited state and ground state is $g_{1}/g_{0} = 2/3$ for $ h \lesssim h_{c1}$ and, respectively, $g_{1}/g_{0} = 3/2$ for $h \gtrsim  h_{c1}$. The height of low-temperature maximum of the specific heat obtained from the Gibbs free energy according to Eq.~(\ref{C_{h}}) is in a good accordance with the prediction of simplified Schottky theory (see Tab.~\ref{piky}). As a matter of fact, both of approaches give the same value $C_{max} / 4k_{\rm{B}} \simeq 0.078 $  and $C_{max} / 4k_{\rm{B}} \simeq 0.153 $ of the specific-heat maximum just below and just above the first critical field $h_{c1} $, respectively [see Fig.~\ref{fig:tetrakap}(a)]. The similar situation can be also found around the second critical field $h_{c2}/J = 3$, where low-temperature thermal excitations between the lowest-energy levels from the sets with the total spin $S_{T}=2$ and $S_{T}=4$ are active. As a consequence of that, one observes the Schottky-type maxima with the height $C_{max} / 4k_{\rm{B}} \simeq 0.032 $ and $C_{max} / 4k_{\rm{B}} \simeq 0.311$ due to the relative degeneracies  $g_{1}/g_{0} = 1/4$ and  $g_{1}/g_{0} = 4$, respectively [see Tab. \ref{piky} and Fig.~\ref{fig:tetrakap}(b)].

All aforementioned results may be independently confirmed by magnetic-field variations of the specific heat plotted in Fig.~\ref{fig:tetrakap}(c). The specific-heat maxima indeed emerge at low enough temperatures just below and above critical fields, whereas their heights are in accordance with the Schottky theory.
The specific heat exactly at critical fields tends to zero on behalf of a lack of low-lying energy levels. This assertion holds for the Ising tetrahedron as well as the other regular Ising polyhedra.

It is important to note that the ground state of any Ising spin cluster is always unique (non-degenerate) at sufficiently high magnetic fields, because all Ising spins are forced to align into the same direction at high enough magnetic fields within the fully polarized state. In this regard, the absolute degeneracy of the regular Ising polyhedra is always known above the saturation field. Only two energy levels cross each other at each critical field of the Ising tetrahedron and hence, one can obtain the absolute degeneracy of ground and first-excited states in a full range of magnetic fields solely from the information about the height of Schottky-type maxima.

\begin{figure*}[t]
\begin{center}
\vspace{-2.0cm}
\hspace{-1.4cm}
\includegraphics[width=0.39\textwidth]{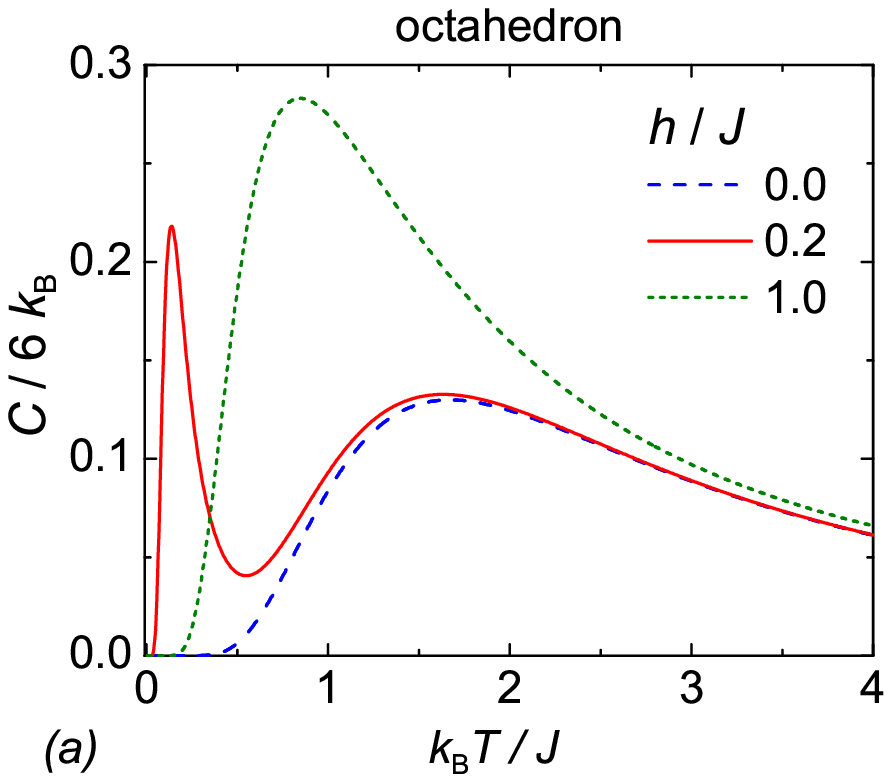}
\hspace{-1cm}
\includegraphics[width=0.39\textwidth]{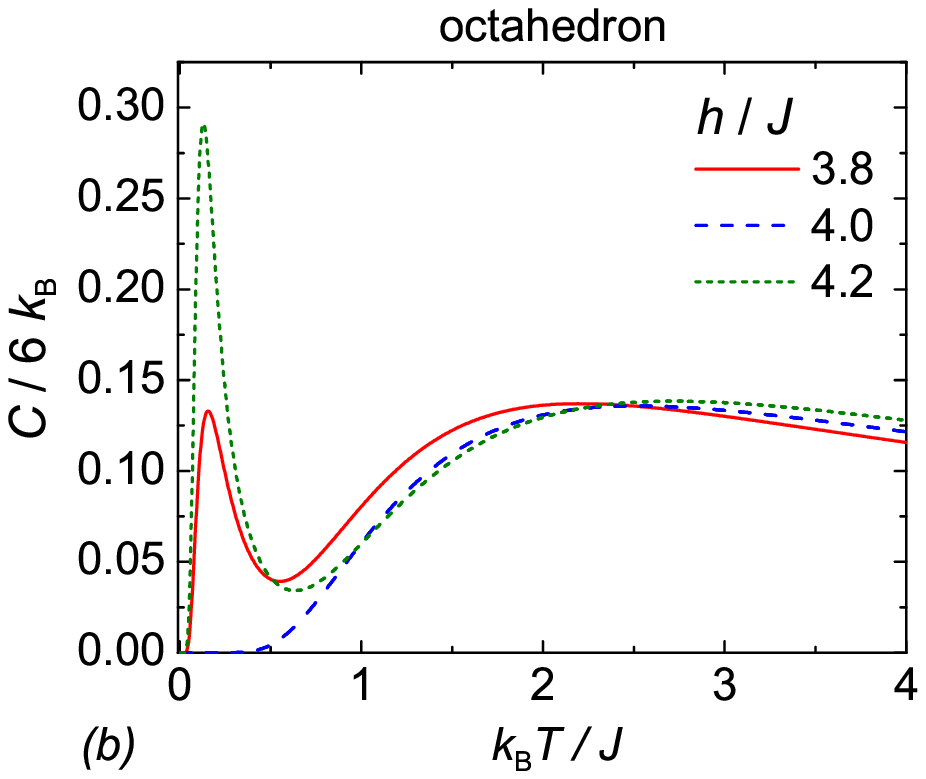}
\hspace{-1cm}
\includegraphics[width=0.39\textwidth]{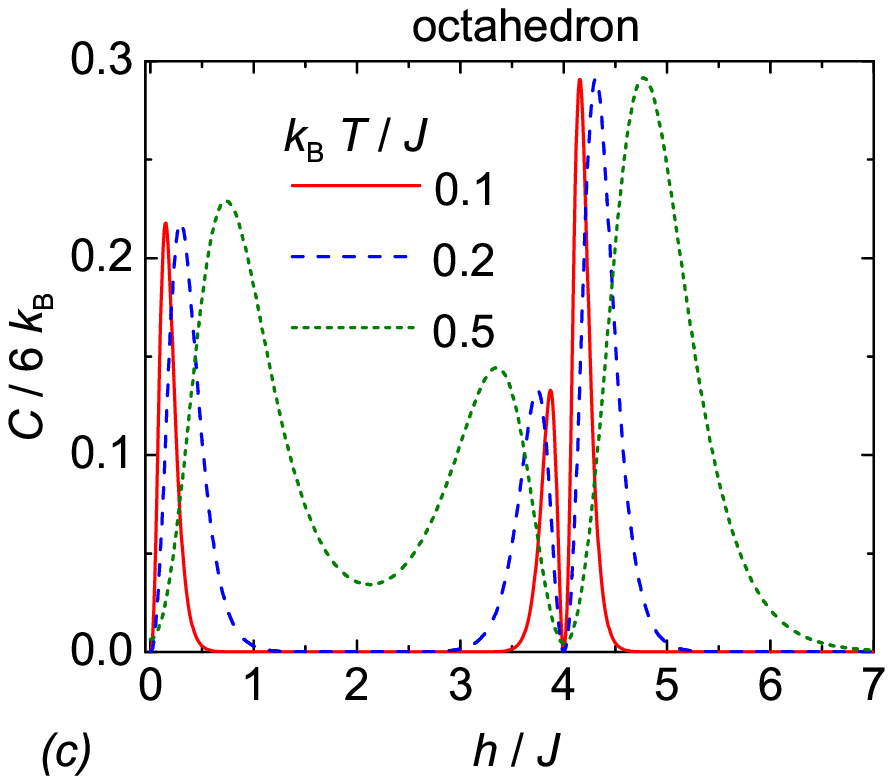}
\hspace{-1.4cm}
\vspace{-0.5cm}
\caption{(a)-(b) Temperature dependences of the specific heat of the Ising octahedron at magnetic fields close to the critical field $h_{c1}=0$ and $h_{c2}=4$; (c) the magnetic-field dependence of the specific heat of the Ising octahedron at different values of temperature.}
\label{fig:octa}
\end{center}
\end{figure*}

The Ising octahedron represents the particular spin cluster, whose lowest-energy levels cross each other already at zero magnetic field [see Fig. \ref{fig:spectrum}(b)]. Therefore, one can observe the low-temperature peak for any small magnetic field above zero critical field $h_{c1}/J = 0$, which is connected with thermal excitations from the lowest-energy level from the set with the total spin $S_{T}=2$ to the lowest-energy level from the set with the total spin $S_{T}=0$. The height of Schottky-type maximum for the corresponding relative degeneracy $g_{1}/g_{0} = 4$ is $C_{max} / 6k_{\rm{B}} \simeq 0.208$ [see Fig.~\ref{fig:octa} (a)]. The low-temperature peak shifts to higher temperatures with increasing of the magnetic field, whereas its width becomes simultaneously broader until it completely merges with a round high-temperature maximum. It was shown in Ref.~\cite{stre15} that the Ising octahedron exhibits a giant magnetocaloric effect at low enough magnetic fields, so it enables an efficient low-temperature refrigeration by the process of adiabatic demagnetization. However, the spin system is useful for low-temperature refrigeration provided its refrigerant capacity is sufficiently high, which requires high enough specific heat at low magnetic fields. The emerging low-temperature maximum of real height $C_{max} / 6k_{\rm{B}} \simeq 0.218$ thus indeed proves that the Ising octahedron represents a sought spin-frustrated structure for real-world refrigerant materials.

The relative degeneracy $g_{1}/g_{0} = 2$ for $h \lesssim h_{c2}$ and $g_{1}/g_{0} = 6$ for  $h \gtrsim  h_{c2}$ relate to the low-lying energy excitations of the Ising octahedron in a vicinity of the second critical (saturation) field $h_{c2} / J = 4$ [see Fig.~\ref{fig:octa}(b)]. It is worth noticing that three energy levels cross each other at the saturation field and thus, one should expect that the Schottky theory will not perfectly coincide with the actual height and position of the low-temperature maximum. The rising temperature causes a gradual overlap of peaks, which can be seen from the field dependence of the specific heat plotted in Fig. \ref{fig:octa}(c) at moderate temperature $k_{\rm{B}}T/J = 0.5$. 

\begin{figure*}
\begin{center}
\vspace{-0.5cm}
\hspace{-1cm}
\includegraphics[width=0.39\textwidth]{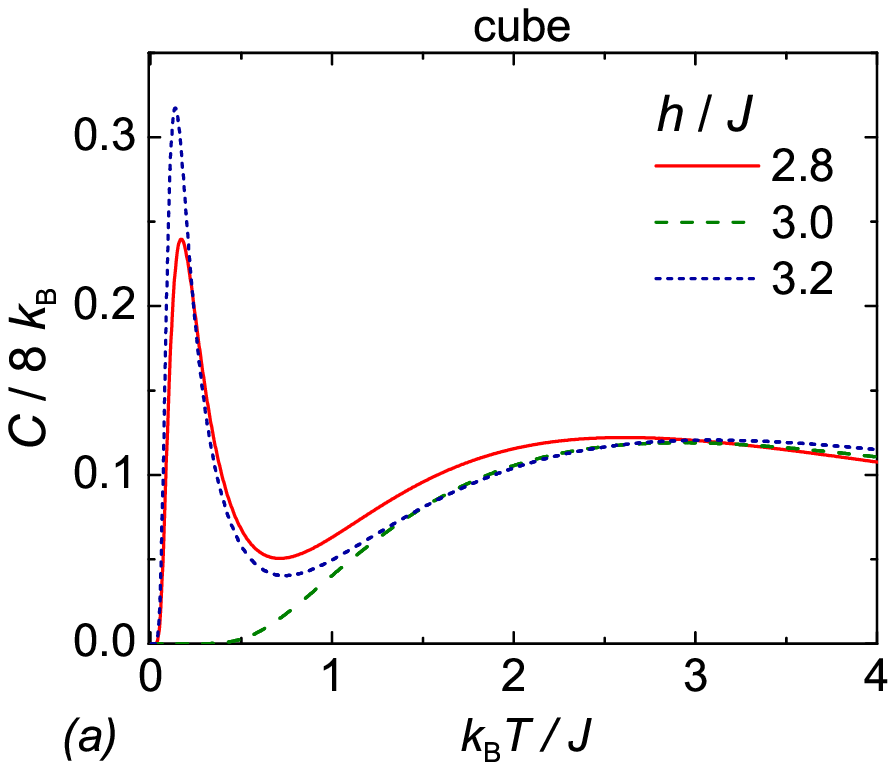}
\hspace{-0.5cm}
\includegraphics[width=0.39\textwidth]{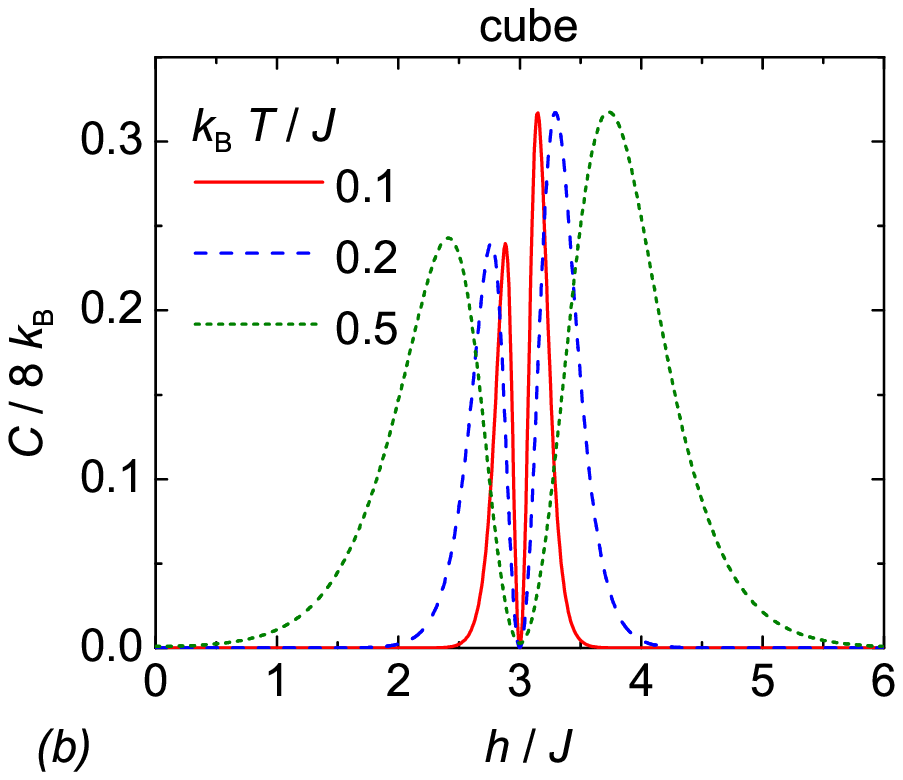}
\hspace{-1.0cm}
\vspace{-0.5cm}
\caption{(a) Temperature dependence of the specific heat of the Ising cube at magnetic fields close to the critical field $h_{c}=3$; (b) the magnetic-field dependence of the specific heat of the Ising cube at different values of temperature.}
\label{fig:cube}
\end{center}
\end{figure*}

The similar effects, which were previously discussed for the specific heat of Ising tetrahedron and octahedron, could be also found in the specific heat of the Ising cube too. The Ising cube is characterized by only one critical field $h_{c}/J = 3$. Low-lying excitations from the lowest-energy level from the set with the total spin  $S_{T}=0$ to the lowest-energy level from the set with the total spin $S_{T}=2$ occurs below this critical field with the appropriate relative degeneracy $g_{1}/g_{0} = 4$, while the higher relative degeneracy $g_{1}/g_{0} = 8$ corresponds to thermal excitation between the lowest-energy levels from the sets with the total spin $S_{T}=6$ and $8$ above this critical field. It can be seen from Fig.~\ref{fig:cube}(a) and (b) that the height of peak at stronger magnetic fields $h \gtrsim  h_{c}$ is higher than the height of peak at weaker magnetic fields $h \lesssim h_{c}$, because the relative degeneracy $g_{1}/g_{0}$ above the saturation field is greater than below it. The Schottky theory of a two-level system describes the specific heat of the Ising cube with much higher inaccuracy in comparison with the Ising tetrahedron or octahedron, because a crossing of four different energy levels takes place at the critical field of the Ising cube [see Fig. \ref{fig:spectrum}(c)]. To be more specific, the real height of low-temperature maximum of the specific heat for $h \lesssim h_{c}$ and $h \gtrsim h_{c}$ is $C_{max} / 8k_{\rm{B}} \simeq 0.234$ and $C_{max} / 8k_{\rm{B}} \simeq 0.317$, while the height of the appropriate Schottky-type maxima would be $C_{max} / 8k_{\rm{B}} \simeq 0.156$ and $C_{max} / 8k_{\rm{B}} \simeq 0.240$, respectively.            

\begin{figure*}
\begin{center}
\vspace{-0.5cm}
\includegraphics[width=0.39\textwidth]{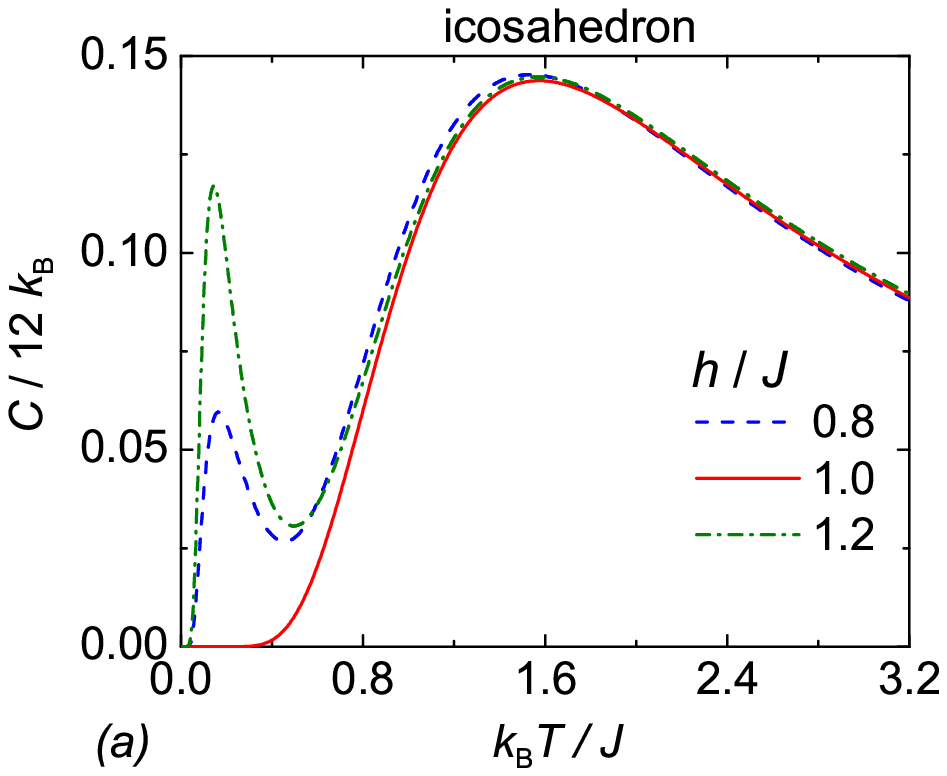}
\hspace{-0.5cm}
\includegraphics[width=0.39\textwidth]{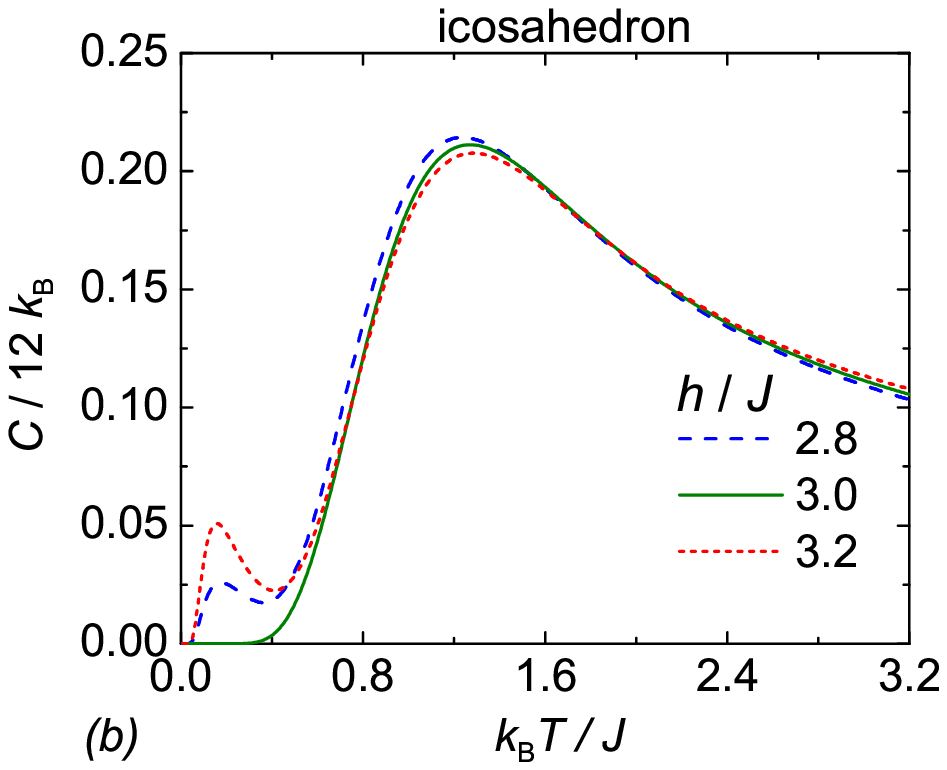} 
\\
\includegraphics[width=0.39\textwidth]{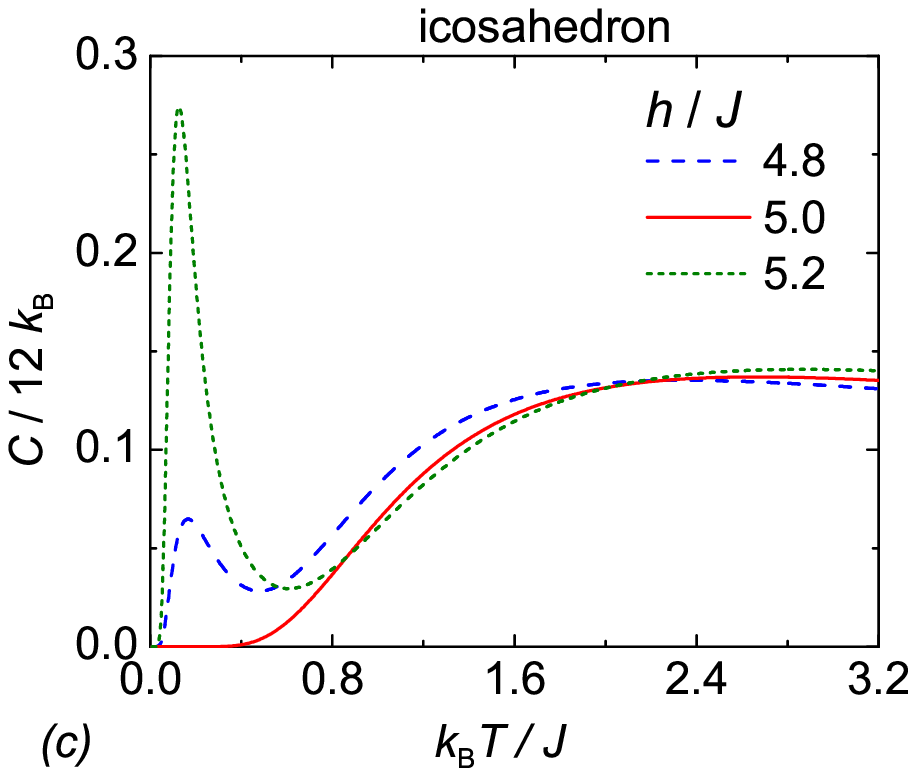}
\hspace{-0.5cm}
\includegraphics[width=0.39\textwidth]{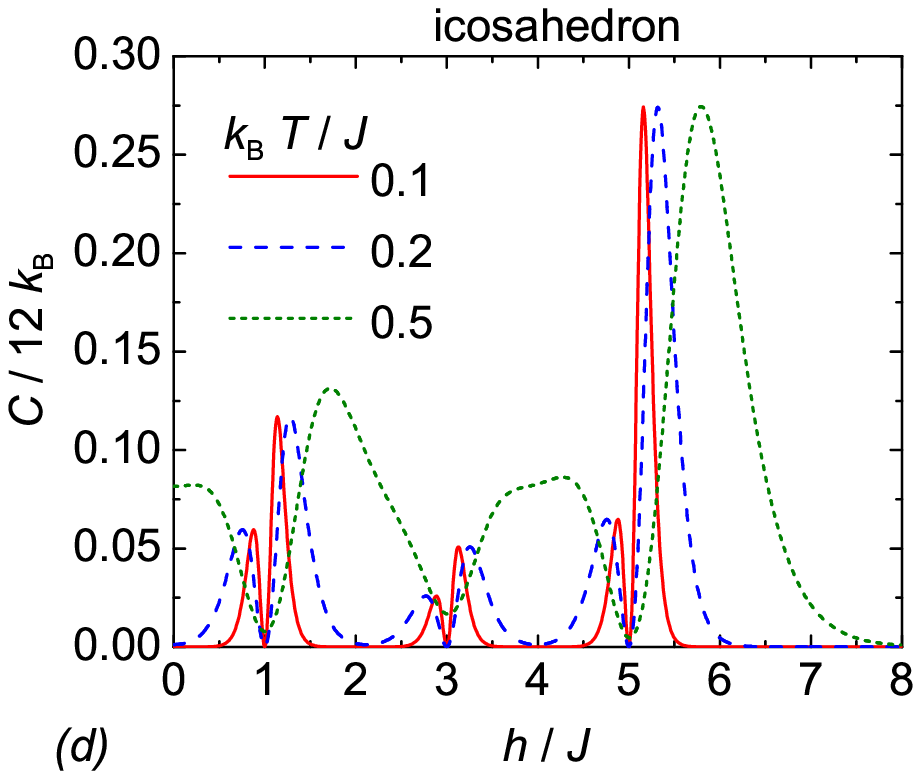}
\vspace{-0.5cm}
\caption{(a)-(c) Temperature dependences of the specific heat of the Ising icosahedron at magnetic fields close to the critical field $h_{c1}=1$, $h_{c2}=3$ and $h_{c3}=5$; (d) the magnetic-field dependence of the specific heat of the Ising icosahedron at different values of temperature.}
\label{fig:icosa}
\end{center}
\end{figure*}

The Ising icosahedron displays the most diverse behavior of the specific heat on account of a more complex energy spectrum [see Fig. \ref{fig:spectrum}(d)], which evidences three different critical fields associated with a crossing of the energy levels. It can bee seen from Fig.~\ref{fig:icosa} that the relevant peaks of the specific heat are always higher when the magnetic field is selected above respective critical field than below it, because the relative degeneracy $g_{1}/g_{0}$ is always greater above the critical fields than below them.

\begin{figure*}
\begin{center}
\vspace{-0.5cm}
\hspace{-1.0cm}
\includegraphics[width=0.39\textwidth]{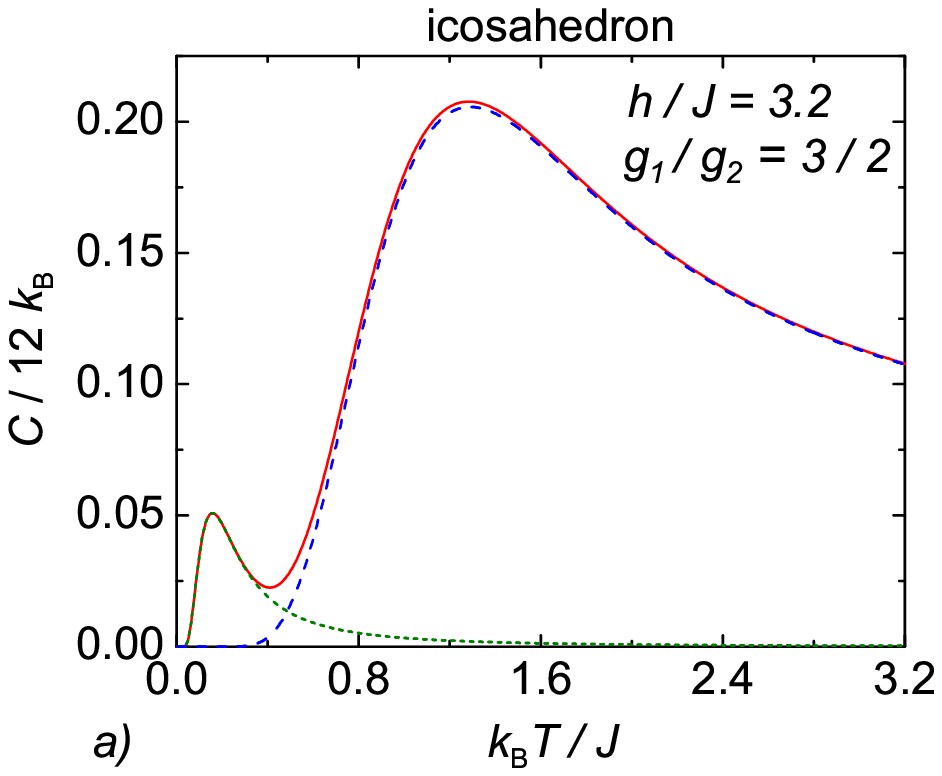}
\hspace{-0.5cm}
\includegraphics[width=0.39\textwidth]{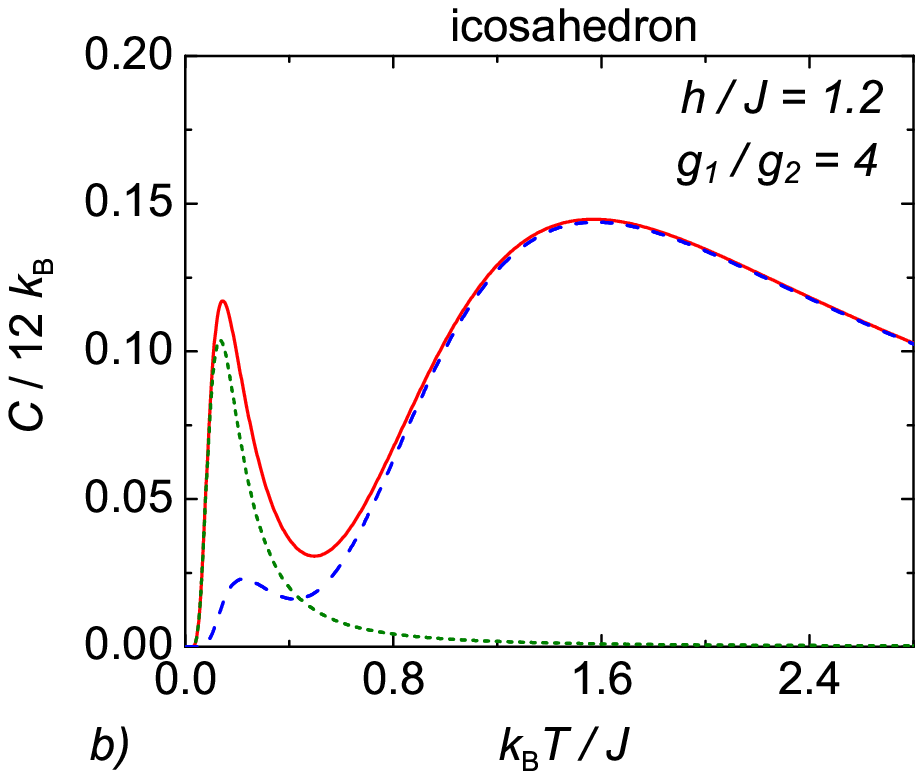} 
\hspace{-1.0cm} 
\vspace{-0.5cm}
\caption{Temperature dependences of the specific heat of the Ising icosahedron calculated from Gibbs free energy (solid curves) and simplified Schottky theory (dotted curves). Dashed curves show their numerical subtraction.}
\label{fig:icosaroziel}
\end{center}
\end{figure*}
A crossing of only two energy levels occurs at the second critical field $h_{c2}/J=3$ of the Ising icosahedron and hence, thermal excitations to any other energy level cannot occur at low enough temperature in a vicinity of the second critical field. It can be seen from Fig.~\ref{fig:icosaroziel}(a) that a perfect conformity between low-temperature dependences of specific heat are obtained from the Schottky theory of a two-level system (dotted curve) and the full model (solid curve). If more than two energy levels cross each other at the critical field, then, the simplified Schottky theory of a two-level system shows a smaller or greater deviations from the real dependence especially as far as the height of the Schottky-type maximum is concerned. The actual height of low-temperature maximum is slightly greater than the height obtained from the simplified Schottky theory due to a presence of thermal excitations to neglected energy levels. An illustrative example of the influence of crossing of more than two energy levels is the specific heat of the Ising icosahedron in a vicinity of the first critical field $h_{c1}/J = 1$ shown in Fig.~\ref{fig:icosaroziel}(b). The subtraction between the exact dependence of the specific heat and Schottky theoretical prediction  implies a small but non-zero deviation even at low-enough temperature due to thermal excitations to one extra  energy level that were neglected [see dashed curve in Fig.~\ref{fig:icosaroziel}(b)].
\begin{figure*}[t]
\begin{center}
\vspace{-0.8cm}
\hspace{-1.4cm}
\includegraphics[width=0.39\textwidth]{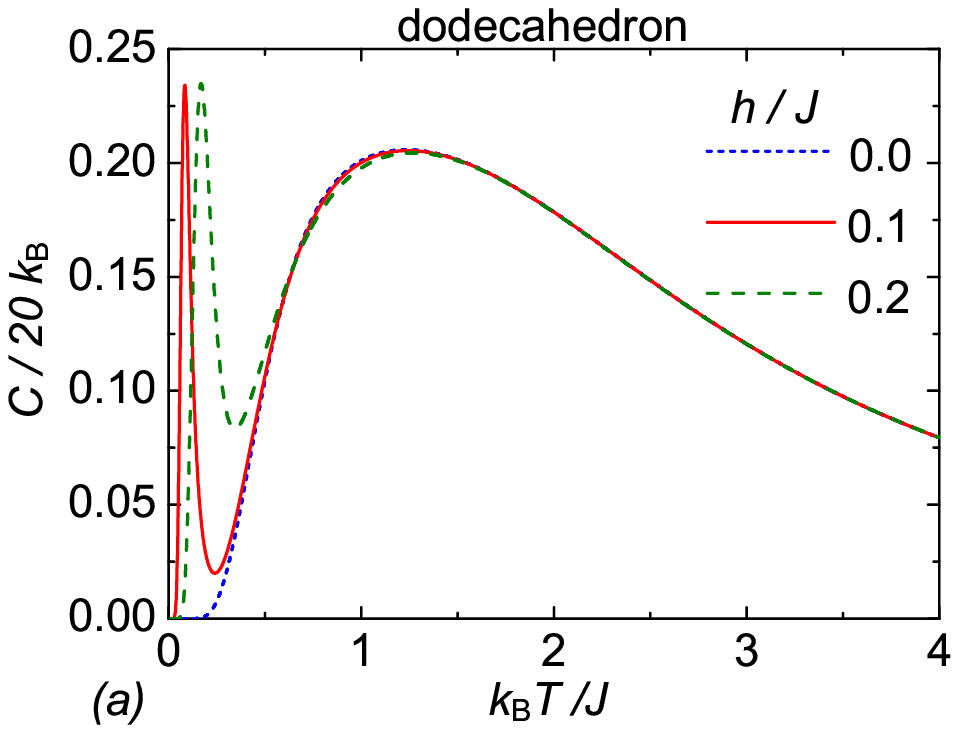}
\hspace{-1cm}
\includegraphics[width=0.39\textwidth]{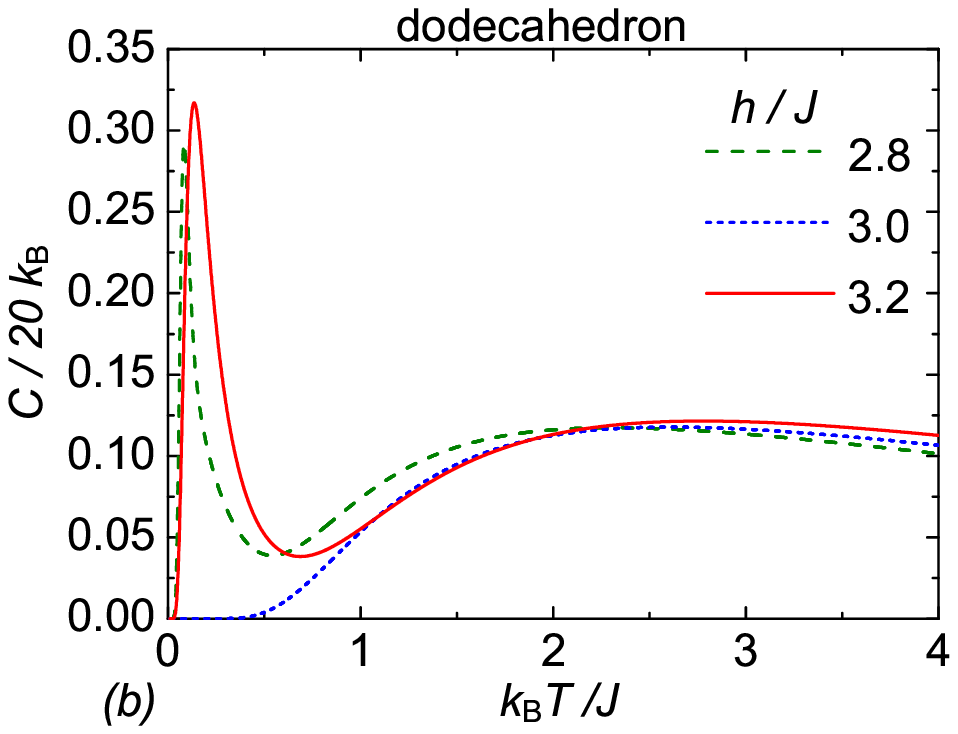}
\hspace{-1cm}
\includegraphics[width=0.39\textwidth]{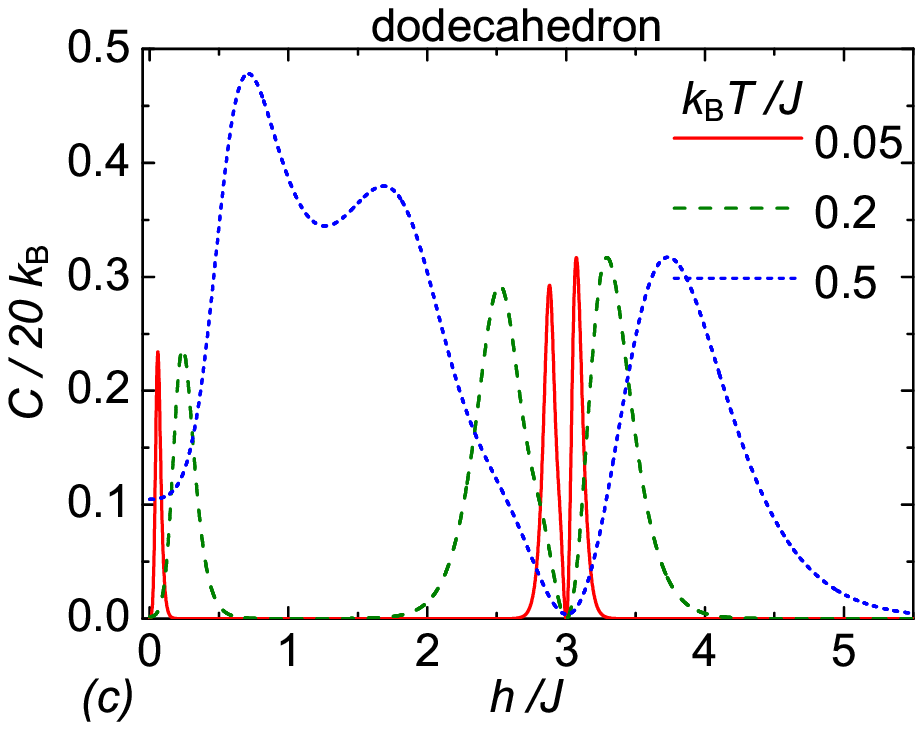}
\hspace{-1.6cm}
\vspace{-0.5cm}
\caption{(a)-(b) Temperature dependences of the specific heat of the Ising dodecahedron at magnetic fields close to the critical field $h_{c1}=0$ and $h_{c2}=3$; (c) the magnetic-field dependence of the specific heat of the Ising dodecahedron at different values of temperature.}
\label{fig:dodeca}
\end{center}
\end{figure*} 

The Ising dodecahedron, similarly as the Ising octahedron, shows the Schottky peaks already at very low magnetic fields due to low-temperature excitations from the ground-state energy level belonging to the set with the total spin $S_{T} = 4$ to the first-excited energy level from the set with the total spin $S_{T} = 0$. The high relative degeneracy $g_{1}/g_{0} = 48$ between both energy levels implies the following height of the Schottky-type maximum $C_{max} / 20k_{\rm{B}} \simeq 0.234$, which is in accordance with of the observed low-temperature maximum of the specific heat of the Ising dodecahedron [see in Fig.~\ref{fig:dodeca}(a)]. As it has been already mentioned at the relevant discussion of the Ising octahedron, the Schottky-type maximum at low enough temperatures indicates 
the applicability of the Ising dodecahedron as a useful refrigerant system, since it exhibits a giant magnetocaloric effect at low magnetic fields quite similarly as does the Ising octahedron \cite{stre15}. The Ising dodecahedron differs from all the other Ising polyhedra by the fact that the preferred low-lying thermal excitation simultaneously flips two Ising spins and hence, the respective change of the total spin equals to $|\delta S|=4$ rather than the more common change $|\delta S|=2$ associated with a single spin-flip thermal excitation \cite{error}. More specifically, the five-fold degenerate ground state of the Ising dodecahedron with the total spin $S_T=4$ corresponds to eight isolated spins flipped in opposite to the magnetic field, while the lowest excited state with the total spin $S_T=0$ consists of four isolated spins and three spin pairs all flipped in opposite to the magnetic field (see the induced subgraphs $8A$ and $10A$ shown in Fig.~2 of Ref. \cite{stre15}).

A crossing of several energy levels occurs at the second critical (saturation) field of the Ising dodecahedron, around which other low-temperature peaks of the specific heat can be detected [see Fig.~\ref{fig:dodeca}(b)]. To be more precise, nine energy levels of the Ising dodecahedron cross each other at the saturation field and thus, one cannot use the simplified Schottky theory of two-level systems as useful approximation in order to determine the height and position of the low-temperature peaks. The field dependence of the specific heat of the Ising dodecahedron is shown in Fig.~\ref{fig:dodeca}(c), which provides independent confirmation of low-lying thermal excitations in a vicinity of zero field. In addition, temperature excitations between energy levels already occur at zero magnetic field as signifies non-zero specific heat at moderate temperatures $k_{\rm B}T/J \approx 0.5$ in a zero-field limit.

To summarize, it has been evidenced that the absolute degeneracy of ground and first-excited states can be determined from the low-temperature maxima of the specific heat of the Ising tetrahedron, while the low-temperature maxima of the other four regular Ising polyhedra (octahedron, cube, icosahedron, dodecahedron) bring insight just into a relative degeneracy at particular level-crossing fields with exception of the saturation field where the absolute degeneracy is available due to the unique character of the fully polarized state.

\section{Conclusion}
\label{conclusion}
The present work deals with the specific heat of the regular Ising polyhedra with the uniform antiferromagnetic nearest-neighbor interaction, which may display striking temperature and field dependences in a vicinity of level-crossing fields. It has been shown that the height and position of observed low-temperature maxima depend basically on a relative degeneracy between the first-excited state and the ground state, whereas the Schottky theory of a two-level system often gives a plausible description of the actual height and position of the specific-heat maxima. In agreement with the Schottky theory, the height of the low-temperature maxima increases and its position decreases with increasing the relative degeneracy between the first-excited state and the ground state. However, the Schottky theory provides a correct quantitative description of the actual height of the low-temperature peaks only if two energy levels cross each other at the relevant critical field, while it underestimates their height when more than two energy levels cross each other due to neglected thermal excitations to other low-lying energy levels. The relative degeneracy between the first-excited amd ground state of the regular Ising polyhedra can be therefore estimated with greater or smaller accuracy from the height and position of low-temperature peaks of the specific heat, whereas the absolute degeneracy is always available at the saturation field with regard to the unique (non-degenerate) character of the fully polarized state. 

It is worthwhile to remark, moreover, that the low-temperature Schottky-type peaks emerging in the specific heat of the Ising octahedron and dodecahedron at low enough magnetic fields indicate an interesting refrigerant capability of these two frustrated spin structures. It has been argued previously \cite{stre15} that a giant magnetocaloric effect accompanies the adiabatic demagnetization of the Ising octahedron and dodecahedron, which consequently undergo a fast cooling at sufficiently low magnetic fields. The relatively high values of the specific heat of the Ising octahedron and dodecahedron $C_{max} / 6k_{\rm{B}} \simeq 0.218$ and $C_{max} / 20k_{\rm{B}} \simeq 0.234$ at low magnetic fields further implies a substantial capacity of magnetic materials based on the Ising octahedron or dodecahedron in order absorb heat from substances that are in thermal contact with them. Thus, it could be concluded that the magnetic materials being composed of a set of non-interacting (or weakly interacting) Ising octahedrons or dodecahedrons can be regarded as promising refrigerant materials. 



\begin{thebibliography}{100}
\bibitem{gopa66} E.S.R. Gopal, \textit{Specific Heats at Low Temperatures}, Heywood Books, London, 1966, pp. 102-105.
\bibitem{path96} K.R. Pathria, \textit{Statistical Mechanics}, Elsevier, Amsterdam, 1996, p. 79.
\bibitem{affr99} M. Affronte, J.C. Lasjaunias, A. Cornia, A. Caneschi, Phys. Rev. B 60 (1999) 1161.
\bibitem{affr02} M. Affronte, J.C. Lasjaunias, G.L. Abbati, Phys. Rev. B 66 (2002) 180405.
\bibitem{luec99} C.S. Lue, J.H. Ross, C.F. Chang, H.D. Yang, Phys Rev B 60 (1999) 941.
\bibitem{tsuj04} H. Tsujii, C. R. Rotundu, Y. Takano, B. Andraka, Y. Aoki, H. Sugawara, H. Sato, J. Magn. Magn. Mater. 272 (2004) 173.
\bibitem{sato14} Y. Sato, H. Morodomi, Y. Inagaki, T. Kawae, H. S. Suzuki, J. Phys.: Conf. Ser. 568 (2014) 042027.
\bibitem{higa02} R. Higashinaka, H. Fukazawa, D. Yanagishima, Y. Maneo, J. Phys. Chem. Solids 63 (2002) 1043.
\bibitem{hama04} N. Hamaguchi, T. Matsushita, N. Wada, Y. Yasui, M. Sato, Phys. Rev. B 69 (2004) 132413. 
\bibitem{fish00} R. A. Fisher, D. A. Wright, J. P. Emerson, B. F. Woodfield, N. E. Phillips, Phys. Rev. B 61 (2000) 538.
\bibitem{popo07} E. A. Popova, D. V. Volkov, A. N. Vasilliev, A. A. Demidov, N. P. Kolmakova, I. A. Gudim, L. N. Bezmaternykh, 
                 N. Tristan, Yu. Skoursky, B. B{\"u}chner, C. Hess, R Klingeler, Phys. Rev. B 75 (2007) 224413.
\bibitem{toko10} N. Tokoro, S. Yamashita, A. Igashira-Kamiyama, J. Fujioka, T.Konno, Y. Nakazawa, J. Therm. Anal. Calorim. 99 (2010) 149.
\bibitem{chak15} T. Chakraborty, H. Singh, Ch. Mitra, J. Magn. Magn. Mater. 396 (2015) 247.
\bibitem{ivan10} N.B. Ivanov, J. Schnack, R. Schnalle, J. Richter, P. K{\"o}gerler, G. N. Newton, L. Cornin, Z. Oshima, H. Nojiri, Phys. Rev. Lett. 105 (2010) 037206.
\bibitem{ozaw05} T.C. Ozawa, T. Taniguchi, Y. Kawaji, Y. Nagata, Y. Noro, H. Samata, S. Takayanagi, Phys. Lett. A 337 (2005) 130.
\bibitem{park00} J.B. Parkinson, J. Timonen, J. Phys.: Condens. Matter 12 (2000) 8669.
\bibitem{kons05} N.P. Konstantinidis, Phys. Rev. B 72 (2005) 064453.
\bibitem{efre06} D.V. Efremov, R.A. Klemm, Phys. Rev. B 74 (2006) 064408.
\bibitem{huch11} A. Hucht, S. Sahoo, S. Sil, P. Entel, Phys. Rev. B 84 (2011) 104438.
\bibitem{schn09a}R. Schnalle, J. Schnack, Phys. Rev. B 79 (2009) 104419.  
\bibitem{schn09b}J. Schnack, R. Schnalle, Polyhedron 28 (2009) 1620.
\bibitem{hone09} A. Honecker, M.E. Zhitomirsky, J. Phys.: Conf. Ser. 145 (2009) 012082.
\bibitem{schn10} J. Schnack, O. Wendland, Eur. Phys. J. B 78 (2010) 535.
\bibitem{stre15} J. Stre\v{c}ka, K. Kar\v{l}ov\'a, T. Madaras, Physica B 466 (2015) 76.
\bibitem{error} Note that the total spin values $S_T=0$, 1 and 2 quoted in Ref. \cite{stre15} in the last paragraph of Section 3 on page 84 
should be replaced with the total spin values $S_T=0$, 2 and 4, respectively, due to the convention $S_i = \pm 1$ used for the Ising spin variables.

\end{thebibliography}
\end{document}